\documentclass[fleqn,12pt,twoside]{article}
\usepackage{espcrc1}

\usepackage{graphicx}
\usepackage[figuresright]{rotating}

\newcommand{\AmS}{{\protect\the\textfont2
  A\kern-.1667em\lower.5ex\hbox{M}\kern-.125emS}}

\title{Few body Calculation of Neutrino Neutral Inelastic scattering on $^4$He}

\author{D. Gazit \address[HUJI]{The Racah Institute of Physics,
        The Hebrew University, Jerusalem, 91904, Israel}
        and
        N. Barnea\addressmark[HUJI]\thanks{This work was supported by
        The ISRAEL SCIENCE FOUNDATION (Grant No. 361/05)}}

\begin{document}

\maketitle

\begin{abstract}
The inelastic neutral reaction of neutrino on $^{4}\mathrm{He}$ is
calculated using two modern nucleon--nucleon potentials. Full final
state interaction among the four nucleons is considered, via the
Lorentz integral transform (LIT) method. The effective interaction
hyperspherical-harmonic (EIHH) approach is used to solve the
resulting {Schr\"{o}dinger} like equations. A detailed energy
dependent calculation is given in the impulse approximation.
\end{abstract}
\section{Introduction}
Neutrino reactions with nuclear targets has led modern physics to
numerous achievements and gave the first hint for deviations from
the standard model. In astrophysics, for example, neutrino
scattering on nuclei are the key ingredient in the supernova
explosion of a massive star and the synthesis of elements within.
The inelastic reactions of $^{4}\mathrm{He}$ with $\nu
_{x}(\overline{\nu _{x}})$ ($x=e,\mu ,\tau $) have a part in these
two phenomena. Core collapse supernovae are widely accepted to be a
neutrino driven explosion of a massive star. When the iron core of a
massive star becomes gravitationally unstable it collapses until
short-range nuclear forces halt the collapse and drive an outgoing
shock through the outer layers of the core and the inner envelope.
However, the shock loses energy through dissociation of iron nuclei
and neutrino radiation, and gradually stalls, it becomes an
accretion shock. Meanwhile, the collapsed core (the proto-neutron
star (PNS)) cools through neutrino radiation, originating in part by
the deleptonization of the core, but mostly by thermally produced
pairs of neutrinos and anti-neutrinos in all flavors. These
neutrinos diffuse out of the PNS. Due to charge current interactions
and electron scattering the electron--neutrinos decouple from matter
in a bigger radius than the heavy-flavored neutrinos. An energy
cascade is thus created, with 6-10 MeV temperature for
$\nu_{\mu,\tau}$ ($\bar{\nu}_{\mu,\tau}$), 5-8 MeV for
$\bar{\nu}_e$, and 3-5 MeV for
$\nu_e$ \cite{WI88}.\\
It is believed that the shock is then revived, as these neutrinos
deposit energy in the matter behind the shock to reverse the flow to
an outgoing shock which explodes the star. This belief was
qualitatively demonstrated in the detection of the neutrino signal
from supernova 1987A \cite{1987A}. However, quantitative proof in
full hydro-reactive simulations is still missing \cite{LI01,BU03}.
The matter in the volume between the PNS and the shock is a hot
dilute gas composed mainly of protons, neutrons, electrons, and
$^{4}\mathrm{He}$ nuclei. In contrast to the fairly known
cross-sections of neutrinos with electrons and nucleons, the
interaction of neutrinos with $^{4}\mathrm{He}$ is not accurately
known. Only recently a first microscopic calculation of the neutral
cross-section has been published \cite{OUR}. \\
The role of $\alpha-\nu$ interaction in the supernova explosion
mechanism is not yet completely understood (see however,
\cite{OH06}). The same interaction, however, has a most important
role in neutrino induced nucleosynthesis \cite{WO90}. The huge
number of neutrinos are the seed of light element nucleosynthesis in
the supernova environment, the so called $\nu$ process. In this
process, an evaporation of a nucleon from $^4$He in the helium rich
layer is followed by a fusion of the trinucleus with another
$\alpha$ nucleus, resulting in a 7 body nucleus. This is an
important source of $^7$Li, and for $^{11}$B and $^{19}$F through
additional $\alpha$ capture reactions. A correct description of this
process must contain an exact, energy dependent cross-section for
the neutral inelastic $\alpha-\nu$ reaction, which initiates the process.\\
In this contribution we give the neutral cross-section in the
impulse approximation, with two different modern NN potentials,
Argonne V8' \cite{AV8P} and V18 \cite{AV18}.\\
\section{Calculation And Results}
The neutrino-nucleus scattering process is governed by the weak
interaction model. In the limit of small momentum transfer (compared
to the Z particle rest mass), the effective Hamiltonian can be
written as a current-current interaction:
$\hat{H}_{W}=\frac{G}{\sqrt{2}}\int {d^{3}xj_{\mu }(\vec{x})J^{\mu
}(\vec{x})}$, where $G$ is the Fermi weak coupling constant, $j_{\mu
}(\vec{x})$ is the leptonic current, and $J^{\mu }$ is the hadronic
current. The matrix element of the leptonic current results only in
kinematical factors to the cross-section. However, the nuclear
current cannot be calculated in such a simple manner. The standard
model dictates only the formal structure:
\begin{equation}
J_{\mu }^{hadronic}=(1-2\cdot \sin ^{2}\theta _{W})\frac{\tau
_{0}}{2}J_{\mu }+\frac{\tau _{0}}{2}{J}_{\mu }^{5}-2\cdot \sin
^{2}\theta _{W}\frac{1}{2 }J_{\mu },  \label{eq:hadcu}
\end{equation}
here we denote axial currents by superscript $5$, and no superscript
denotes the vector currents. The latter are built of both isoscalar
and isovector parts, whereas the axial currents are pure isovector
operators. The nuclear current matrix elements consists of one body
weak currents, but also many body corrections due to meson exchange.
The many-body currents are a result of meson exchanged between the
nucleons. The current work is done in the impulse approximation,
thus taking into account only one-body terms. In order to estimate
this approximation, we refer to studies of inclusive electron
scattering off $^{4}\mathrm{He}$ \cite{CA02}, where it is shown that
isovector electromagnetic two-body currents, which are proportional
to the electroweak vector currents, produce a strong enhancement of
the transverse response at low and intermediate energies. In the
current calculation, the vector MEC are fully considered within the
Siegert theorem, and the two-body axial currents are expected to
give small contributions \cite{SC03,OURPRL}.\\
The one-body currents connect the $^{4}\mathrm{He}$ ground state and
final state wave functions. An explicit calculation of the nuclear
response functions demands an accurate description of all the
excited states of the nucleus. For $^4$He this is currently out of
reach, as the nucleus has no excited bound states, and no accurate
continuum wave functions exist for all channels of system with more
than 3 particles. In order to avoid this complexity we use the
Lorentz integral transforms \cite{EF94}. As a result, the problem
reduces to a bound state like problem. A most successful method for
solving these kind of problems is the EIHH \cite{BA00} method. The
combination of the EIHH and LIT methods brings to a rapid
convergence in the response functions \cite{OUR,GA06}.\\
It is of interest to note that since $^4$He is almost pure isoscalar
and approximately spin zero, the leading contributions to the
cross-section are proportional to $qr$, where $q$ is the momentum
transfer and $r$ is the coordinate of the nucleon. The contribution
of the isoscalar current to the neutral cross-section is of higher
order in
this small parameter, thus negligible.\\
\begin{table}
\caption{{\label{tab:crs}} Flavor and temperature averaged inclusive
inelastic cross-section}
\begin{tabular}{cccc}
\hline T [MeV] &  \multicolumn{3}{c}{$\langle \sigma \rangle_T$
[$10^{-42}cm^{2}$] }  \\
 & \hspace{0.15cm} AV8' & \hspace{0.25cm} AV18
 & \hspace{0.5cm} Ref.~\cite{WO90}\\
\hline
 4    &  2.09(-3) & 2.31(-3) &    -     \\
 6    &  3.84(-2) & 4.30(-2) & 3.87(-2) \\
 8    &  2.25(-1) & 2.52(-1) & 2.14(-1) \\
 10   &  7.85(-1) & 8.81(-1) & 6.78(-1) \\
 12   &  2.05     & 2.29     & 1.63     \\
 14   &  4.45     & 4.53     &    -     \\
 16   &  8.52     & 9.48     &    -     \\
\hline
\end{tabular}
\end{table}
It is assumed that the neutrinos are in thermal equilibrium, thus
their spectrum can be approximated by the Fermi-Dirac distribution
with characteristic temperature $T$. As a result, the most
interesting physical quantities are the temperature averaged
cross-section, ${\frac{d\langle\sigma\rangle_T}{d\omega}=\int dk_i
f(T,k_i)\frac{d\sigma}{dk_f}}$, and energy transfer cross-section,
${\frac{d\langle\sigma \omega \rangle_T}{d\omega}=\omega
\frac{d\langle\sigma\rangle_T}{d\omega}}$, where $f(T,k)$ is
normalized Fermi-Dirac spectrum with zero chemical potential,
temperature $T$, and energy $k$. In Table~\ref{tab:crs} we present
the calculated total temperature averaged cross-section,
$\langle\sigma\rangle_T=\frac{1}{2} \frac{1}{A} \langle
\sigma_\nu+\sigma_{\overline{\nu}}\rangle_T$, as a function of the
temperature of the neutrinos. Also presented are earlier results
reported by Woosley et. al. \cite{WO90,HA88}. It can be seen that
the current work predicts a substantial enhancement in the
cross-section. \\
The energy transfer cross-section was fitted by Haxton to a
convenient formula \cite{HA88},
\begin{equation}
\langle\sigma \omega \rangle_T = \alpha \left( \frac{T-T_0}{10
\rm{MeV}}\right)^\beta
\end{equation}
with the parameters $\alpha=0.62 \cdot \rm{10^{-40} cm^2 MeV}$,
$T_0=2.54 \rm {MeV}$, $\beta=3.82$. A similar fit to our results
yields for AV8': $\alpha=0.64 \cdot \rm{10^{-40} cm^2 MeV}$, $T_0=
2.05 \rm {MeV}$, $\beta=4.46$, and for AV18: $\alpha=0.72 \cdot
\rm{10^{-40} cm^2 MeV}$, $T_0= 2.12 \rm {MeV}$, $\beta=4.42$. It can
be seen that the current work predicts a stronger temperature
dependence of the cross sections. For example, a $15-30\%$ differnce
between these calculations at $T=10$ Mev, grows to a $50-70\%$
difference at $T=16$ MeV.
\section{Conclusions}
A detailed realistic calculation of the inelastic
neutrino-$^{4}\mathrm{He}$ neutral scattering cross-section is
given. The calculation was done in the impulse approximation with
numerical accuracy of about $1\%$. The different approximations used
here should result in about $10\%$ error, mainly due to many-body
currents and three nucleon force (3NF), which were not considered in
the current work. Axial vector part contributes more than $90\%$ of
the cross-section given. As these currents are not protected by
current conservation, they should be included explicitly in the
calculation. However, comparing to the size of these effects in
electron scattering processes, one should expect only a few percents
effect. \\
The nuclear hamiltonian used in the current work consists only of
nucleon-nucleon potential, and neglects 3NF. Lately, a detailed
description of the photoabsorption on $^4$He has shown that
including 3NF results in a reduction of $10\%$ of the cross-section
\cite{GA06}. Thus, one should also check this effect on
neutrino-$^4$He
cross-section. \\
The influence of the current calculation on the supernova explosion
mechanism should be checked through hydrodynamic simulations, of
various progenitors. Nonetheless, it is clear that our results
facilitate a stronger neutrino-matter coupling in the supernova
environment. First, our calculations predict an enhanced cross
section with respect to previous estimates. Second, we obtained
steeper dependence of the energy transfer cross-section on the
neutrino's temperature. Thus, supporting the observation that the
core temperature is a critical parameter in the explosion process.
It is important to notice that the energy-transfer due to inelastic
reactions are $1-2$ orders of magnitude larger than the elastic
reactions, ergo the inelastic cross-section are important to an
accurate description of the Helium shell temperature.

\end{document}